\documentclass[pra,twocolumn,showpacs,amssymb,superscriptaddress,nofootinbib]{revtex4}
\usepackage[utf8x]{inputenc}
\usepackage{epstopdf}
\usepackage[colorlinks=true, citecolor=blue, urlcolor=blue]{hyperref}
\usepackage{amsmath}
\usepackage{graphicx}
\usepackage{dcolumn}
\usepackage{bm}
\usepackage{color}
\usepackage{longtable}
\usepackage{amsthm}

\newtheorem*{theorem*}{Theorem}

\newtheorem{proposition}{Proposition}

\begin{document}

 \title{Inferred-variance uncertainty relations in the presence of quantum entanglement}

\author{Shrobona Bagchi}
\email{shrobonab@mail.tau.ac.il}
\affiliation{Raymond and Beverly Sackler School of Physics and Astronomy, Tel Aviv University, Tel-Aviv 69978, Israel.}

\author{Chandan Datta} 
\email{c.datta@cent.uw.edu.pl}
\affiliation{Centre for Quantum Optical Technologies, Centre of New Technologies, University of Warsaw, Banacha 2c, 02-097 Warsaw, Poland.}

\author{Pankaj Agrawal}
\email{agrawal@iopb.res.in}
\affiliation{Institute of Physics, Sachivalaya Marg,
Bhubaneswar 751005, Odisha, India.}
\affiliation{Homi Bhabha National Institute, Training School Complex, Anushakti Nagar, Mumbai 400085, India.}

\begin{abstract}
Uncertainty relations play a significant role in drawing a line between classical physics and quantum physics. 
Since the introduction by Heisenberg, these relations have been considerably explored.
However, the effect of quantum entanglement on uncertainty relations was not probed. 
Berta et al. [Nature Physics \textbf{6}, 659-662 (2010)] removed this gap by deriving a conditional-entropic uncertainty relation in the presence of quantum entanglement. In the same spirit, using inferred-variance, we
formulate uncertainty relations in the presence of entanglement for general two-qubit systems and arbitrary observables. We derive lower bounds for the sum and
product inferred-variance uncertainty relations. Strikingly, we can write the lower bounds of
these inferred-variance uncertainty relations in terms of measures of entanglement of two-qubit states, as characterized by concurrence, or $G$ function. Presumably, the presence of entanglement in the lower
bound of inferred-variance uncertainty relation is new and unique. We also explore the violation of local uncertainty relations in this context and an interference experiment. Furthermore, we discuss possible applications of these uncertainty relations.

\end{abstract}

\maketitle

\section {Introduction }\label{intro}
In quantum theory, linear superposition between different quantum states gives rise to the phenomenon of quantum 
interference and uncertainty relations in Hilbert space \cite{H,rob,sch}. The field of uncertainty relations has grown significantly since the time of Heisenberg and even has proved to be crucial in the implementation of various quantum information and computational tasks \cite{d,berta}.
On the other hand, superposition also leads to the phenomenon of entanglement in bipartite quantum systems. 
Quantum entanglement is behind many intriguing features of quantum mechanics. It is also a useful resource for many 
communication and computational tasks \cite{horodecki}. Although they both feature the same phenomenon of superposition, 
noticeably the presence of entanglement reduces uncertainty between two non-commuting observables. In this article, 
we are interested in exploring how inferred-variance uncertainty relations are affected by the entanglement in a bipartite quantum system.

In the field of uncertainty relation, there exist two types of uncertainty relations, namely the preparation uncertainty relations and the measurement-disturbance relations that capture the uncertainty for
non-commuting observables \cite{H,rob,sch,Lahti,Busch1}. 
After the introduction by Heisenberg \cite{H}, various kind of uncertainty relations were introduced and shown to be useful in various quantum information and quantum computation tasks \cite{coles}--Robertson type \cite{rob,sch,M}, uncertainty relations based on R\'{e}nyi entropy, Tsallis entropy \cite{coles}, conditional entropy \cite{maassen}, and two-state vector formalism \cite{Vaidman1, Vaidman2} to name a few. Despite these developments, uncertainty relations modified by or showing the presence of entanglement were missing. In 1989, Reid introduced a variance based uncertainty relation whose violation confirms presence of entanglement in the state \cite{reid}. However, a lower bound explicitly composed of an entanglement term was missing until recently. In 2010, Berta \textit{et al.} showed that the lower bounds of conditional entropic uncertainty relations depend on the entanglement of bipartite quantum states \cite{berta}.

Uncertainty relations in the presence of quantum entanglement as formulated for entropic uncertainty relation in \cite{berta}, consist of the entanglement signature given by the conditional von-Neumann entropy, 
quantum discord as well as classical correlation \cite{pati}. These correlation measures contribute to the reduction of conditional entropic uncertainty \cite{berta,pati}. However, variance based uncertainty relations in the presence of quantum entanglement have still not been formulated. 
Keeping in view that entropic conditional uncertainty relations have found use in the quantum information and communication tasks, 
and that normal variance uncertainty relations also have some important applications, we are motivated to formulate uncertainty relations using inferred variances, 
which also can result in important applications in quantum information processing tasks in the future. On the top of that, it is noteworthy to mention that the variance can be measured quite easily in an experiment, which is another motivation to look for 
variance based uncertainty relations.

In this paper, we formulate inferred-variance uncertainty relations for 
correlated two-qubit
quantum systems. We find
that the lower bound of inferred-variance uncertainty relations for 
two non-commuting Pauli observables can be connected to the entanglement of the bipartite system. Consequently, as variance based quantities can be easily measured in an experiment, we can put a lower bound on the entanglement of the state in terms of variances. We also discuss possible applications of the derived inferred-variance uncertainty relations in measuring the entanglement of a pure bipartite state
and determining the security of a quantum cryptography protocol.

The paper is organized as follows. In Section \ref{background}, we review the necessary background and results in the literature that we use in our analysis. We present our main results relating to the lower bound of inferred-variance uncertainty relations with the entanglement of bipartite two-qubit systems in Section \ref{result}.
In Section \ref{local uncertainty}, we 
discuss the role of entanglement in the violation of local 
uncertainty relations. 
Along the way, we find relations for local uncertainty relations that also involve entanglement. Possible applications of inferred-variance uncertainty relations have been discussed qualitatively in Section \ref{application}. In Section  \ref{visibility}, 
we put this in the context of an interference experiment where the change in the visibility of the interference fringes is constrained by the entanglement measure of a quantum state. Finally,
in Section \ref{conclusion}, we state our conclusions and point out the future directions.

\section {Background }\label{background}

Here in this section we briefly discuss some necessary background.
The topics needed for our analysis include entropic uncertainty relations in the presence of quantum memory \cite{berta,pati}, 
inferred uncertainty relations \cite{reid}, violation of local uncertainty relations \cite{hoffman}, entanglement and quantum correlation measures \cite{horodecki}, 
and experimental set up related to the interference fringe visibility connected to the unitary operators acting on one arm of the interferometer \cite{bagchi}.

\subsection {Uncertainty relations with quantum entanglement}

The original formulation of uncertainty relation was for a single
quantum system. It was soon realized that this relation can be violated in the presence of entanglement, however the exact quantitative equation was not formulated.
Later, after being cast in the form of a quantum game, Berta {\ it et al.} in \cite{berta} showed that the lower bound of the uncertainties about the outcomes of two incompatible measurements can be further reduced in the presence of quantum entanglement. The lower bound in their work depends on the entanglement present in the system. Mathematically it is represented as follows
\begin{equation}\label{berta bound}
H(R|B)+H(S|B)\geq \log_2\frac{1}{c}+H(A|B),
\end{equation}
where $R$ and $S$ represent some observables. Here $c=\max_{i,j}|\langle \psi_i|\phi_j\rangle|^2$, with $|\psi_i\rangle$ and $|\phi_j\rangle$ represent the eigenvectors of the observables $R$ and $S$ respectively. The uncertainty in the measurement outcome is represented by the conditional von-Neumann entropy $H(\cdot|B)$ with respect to $B$ and $H(A|B)$ represents the amount of entanglement between $A$ and $B$.
Can a similar situation, if not exactly the same, be thought of in terms of inferred variance? For this, we next describe the inferred uncertainty relations \cite{reid}.

\subsection {Inferred uncertainty relations}

A version of uncertainty relation that takes into account the inferred variance is known as inferred uncertainty relation.  \cite{reid}. In this set up also, there are two different
experimenters Alice and Bob. 
Let us consider two 
spatially separated subsystems $A$, $B$, and two observables that do not commute with each other. When these subsystems are entangled,
then one can predict the result of 
measurement on $A$ based on the result of measurement on $B$. Here, the main task is to what extent we can predict the result of measurement on $A$ based on the result of measurement 
on $B$, for two non-commuting observables, using the uncertainty quantifier in terms of variance. In this definition, it has been shown that the violation of the following uncertainty relation is a signature of the EPR steering criteria \cite{reid}
\begin{equation}\label{reid_uncertainty}
\Delta^2 S_{\inf}\Delta^2 Q_{\inf}\geq L,
\end{equation}
where $S$ and $Q$ are two non-commuting observables and $L$ is the lower bound for this set of observables. 
Since all the states which are steerable are also entangled, 
the violation of the above uncertainty relation is also a signature of entanglement. Indeed, it was shown in 
\cite{reid, cavalcanti} that the violation of inferred uncertainty relations can also be used as an entanglement detection criteria. 

\subsubsection{Recipe to calculate inferred variance:} 

We follow the recipe proposed in \cite{reid} to find out the inferred variances.
Consider a two-qubit state $\rho_{AB}$. We want to infer the variance of the general Pauli measurement $S$ 
on Alice's side from the measurement outcome of $S$ on Bob's side. We denote the inferred outcome on Alice's side by $i=0,1$ and the measurement outcome on Bob's side 
by $j=0,1$. The recipe is as follows:
The probability of getting outcome $i$ and $j$ is
\begin{equation}
P(i,j)=\mbox{Tr}[|i\rangle\langle i|\otimes |j\rangle\langle j|\rho_{AB}],
\end{equation}
where $|i\rangle\langle i|$ denotes the projector for $i$th outcome. Summing over the outcomes of Alice, we will find the probability of getting outcome $j$ on Bob's side
\begin{equation}
P(j)=\sum_{i}P(i,j).
\end{equation}
Now to find out the conditional probability of obtaining outcome at $A$ conditioned on the measurement outcome at $B$, we use the following
\begin{equation}
P(i|j)=\frac{P(i,j)}{P(j)}.
\end{equation}
The conditional mean value on Alice's side based on the conditional probability distribution defined above is given by
\begin{equation}
\mu_j=\sum_i x_iP(i|j),
\end{equation}
where $x_0=1$ and $x_1=-1$ denote the original value of the outcome on Alice's side. So, by predicting the outcome on Alice's side how much error we made can be calculated as
\begin{equation}\label{inferred variance}
\Delta^2 S(\rho_{AB})_{\inf}=\sum_{i,j}P(i,j)\left(x_i-\mu_j\right)^2,
\end{equation}
which is the inferred variance \cite{reid}.

\subsection {Violation of local uncertainty relations}

There are various methods to detect the presence of entanglement in a quantum state. One such approach was formulated by Hoffman et al \cite{hoffman}, 
by quantifying the violation of local 
uncertainty relations with respect to a global one. 
Suppose we have a quantum state $\rho_{AB}$, whose reduced density matrices are $\rho_A$ and $\rho_B$. The individual quantum 
systems $\rho_A$ and $\rho_B$ satisfy the usual uncertainty relation bounds which may be state dependent or state independent. Therefore, if we have a set of non-commuting observables
$\{A_i\}$ on side $A$ and $\{ B_i\}$ on side $B$, then the local uncertainty relations are
\begin{equation}
\sum_i\Delta^2 A_i\geq L_A ~~~ \mbox{and}    ~~~\sum_i\Delta^2 B_i\geq L_B,
\end{equation}
where $\Delta^2 A_i=\mbox{Tr}(A_i^2\rho_A)-[\mbox{Tr}(A_i\rho_A)]^2$ is the usual variance uncertainty quantifier, similar for $B_i$ as well. $L_A$ and $L_B$ are the state 
independent lower bounds. 
Now, if we introduce another quantity $\Delta^2(A_i+B_i)$, defined in the same way as shown in \cite{hoffman}, then the following inequality is satisfied only when 
$\rho_{AB}$ is a separable state
\begin{equation}\label{local uncertainty relation violation}
\sum_i\Delta^2 (A_i+B_i)\geq \sum_i\Delta^2 A_i +\sum_i\Delta^2 B_i\geq L_A+L_B.
\end{equation}
It was shown in \cite{hoffman}, that any violation of the above inequality is a definitive signature of quantum entanglement. 

\subsection {Connected correlator and its relation with entanglement measure/witness }

\subsubsection{Connected correlator}

For a bipartite density matrix $\rho_{AB}$ in $H_A\otimes H_B$, the bipartite connected correlator is given by the following function
\begin{equation}
CC(A,B)=\mbox{Tr}[(A\otimes B)\rho_{AB}]-\mbox{Tr}[A\rho_{A}]\mbox{Tr}[B\rho_{B}],
\end{equation}
where $A$ and $B$ are the two observables defined in the Hilbert spaces $H_A$ and $H_B$.
There are interesting connections between the connected correlation function and other entanglement measures. At the most fundamental level, 
a non-zero value of the connected correlation function for a bipartite pure state signifies the presence of entanglement in the quantum state as has been shown in \cite{tran}. 
However, in the case of a mixed state, the answer is not definite as a separable state can also show a nonzero value \cite{1, dagomir}.  Below we briefly describe the connections
of connected correlator with concurrence for pure two-qubit states and G function for two-qubit mixed states.

\subsubsection{Connection with entanglement measure/witness}

\subsubsection*{Pure states}
At first, we discuss the relation of the connected correlator with the concurrence. It has been shown in \cite{Popp} that the entanglement as measured by the concurrence of 
a two-qubit pure state is given by the connected correlation function maximized over all the sets of arbitrary Pauli observables. This is represented in the following way
\begin{equation}\label{concurrence and cc}
C(|\psi_{AB}\rangle)=\max_{\vec{a},\vec{b}}|CC({\sigma_{\vec{a}},\sigma_{\vec{b}}})|,
\end{equation}
where $\vec{a}$ and $\vec{b}$ represent the Bloch vector in $\mathbb{R}^3$ and the arbitrary Pauli observable can be constructed as $\sigma_{\vec{n}}=\vec{n}.\vec{\sigma}$.

\subsubsection*{Mixed states}

There is an entanglement witness based on the connected correlation function of local observables that can be applied for the case of two-qubit mixed states. 
This was explored in \cite{1,2,3}.
In particular it is given by 
\begin{equation}
G(\rho_{AB})=\sum_{i,j=1}^3 CC(\sigma_i,\sigma_j)^2,
\end{equation}
where $\sigma_i$ corresponds to the Pauli operators. Note that $G(\rho_{AB})\geq 1$, states that the state $\rho_{AB}$ is entangled.

\section{Inferred uncertainty relations bounded by entanglement for two qubits and the Pauli observables} \label{result}

In this section, we derive a few inequalities that show how entanglement appears in the lower bound of inferred-variance uncertainty relations. First we describe the steps to calculate the inferred variance.

\subsection{Results}

\subsubsection{Sum inferred-variance uncertainty relations}

\begin{theorem*} 
For any two arbitrary Pauli observables $S=\vec{n}.\vec{\sigma}$ and $Q=\vec{m}.\vec{\sigma}$ and an arbitrary two-qubit density matrix $\rho_{AB}$, the 
following equality holds
\begin{eqnarray}
&&\Delta^2 S(\rho_{AB})_{\inf}+\Delta^2 Q(\rho_{AB})_{\inf}= \Delta^2 S_A+\Delta^2 Q_A-\nonumber\\ 
&&\Big[\frac{CC(S,S)^2}{\Delta^2 S_B}+\frac{CC(Q,Q)^2}{\Delta^2 Q_B}\Big],\label{correlator}
\end{eqnarray}
where $CC(S,S)$ and $CC(Q,Q)$ denote the connected correlators of the state $\rho_{AB}$ for observables $S$ and $Q$ respectively.
The first two uncertainties on the right hand side denote local uncertainties of the observables in party $A$, the last two are that of party $B$.
\end{theorem*}
\begin{proof} 
 Let a general two-qubit density matrix be written as the following
\begin{equation}
\rho_{AB}=\frac{1}{4}\Big[I+\sum_{i=1}^3 r_i \sigma_{i}\otimes I +\sum_{i=1}^3s_i I\otimes\sigma_i+\sum_{i,j=1}^3 t_{ij}\sigma_i\otimes \sigma_j\Big].
\end{equation}
To make expressions simpler let us denote $\vec{n}.\vec{r}=nr$, $\vec{n}.\vec{s}=ns$ and $\sum_{i,j=1}^3{n_in_jt_{ij}}=ntn$. According to this notation we get the following quantities. The local uncertainties in party $A$ and $B$ due to Pauli observable $\vec{n}.\vec{\sigma}$ and $\vec{m}.\vec{\sigma}$ are given by 
\begin{eqnarray}
&&\Delta^2 S_A=1-(nr)^2; \Delta^2 S_B=1-(ns)^2\,\, \mbox{and} \label{lounS}\\
&&\Delta^2 Q_A=1-(mr)^2; \Delta^2 Q_B=1-(ms)^2 \label{lounQ}
\end{eqnarray}
respectively. The expressions of the connected correlators $CC(S,S)$ and $CC(Q,Q)$ when we have $\vec{n}.\vec{\sigma}$ and $\vec{m}.\vec{\sigma}$ acting on both the parties $A$ and $B$ are given by 
\begin{eqnarray}
&&CC(S,S)=ntn-(nr)(ns)\,\, \mbox{and}\label{ccnn}\\
&&CC(Q,Q)=mtm-(mr)(ms)\label{ccmm}
\end{eqnarray}
respectively. The initial expression of variance based inferred uncertainty $\Delta^2 S(\rho_{AB})_{\inf}$ is a non-trivial one, and is given in Appendix \ref{expression of inferred}. Carefully rearranging the terms and doing some algebra, we find the following expression for $\Delta^2 S(\rho_{AB})_{\inf}$ for an observable $\vec{n}.\vec{\sigma}$
\begin{eqnarray}\label{inferred varianceS}
&&\Delta^2 S(\rho_{AB})_{\inf}=\\
&&\frac{1}{2}\Big[\frac{(1-ns)(1-nr-ns+ntn)(1-ns-ntn+nr)}{(1-ns)^2}\nonumber \\
&&+\frac{(1+ns)(1-nr+ns-ntn)(1+ns+ntn+nr)}{(1+ns)^2}\Big].\nonumber
\end{eqnarray}
The quantity $\Delta^2 S(\rho_{AB})_{\inf}$ is calculated using the recipe provided above as in Eq. (\ref{inferred variance}). Simplifying the above expression we get the following
\begin{eqnarray}
\Delta^2 S(\rho_{AB})_{\inf}&=&\frac{1}{[1-(ns)^2]}\big[1-(nr)^2-(ns)^2-\nonumber\\
&&(ntn)^2+2(nr)(ns)(ntn)\big]\nonumber \\
&=&[1-(nr)^2]-\frac{[ntn-(nr)(ns)]^2}{[1-(ns)^2]}\nonumber\\
&=&\Delta^2 S_A-\frac{CC(S,S)^2}{\Delta^2 S_B},\label{infS}
\end{eqnarray}
where in the last step we use the relations in Eqs. (\ref{lounS}) and (\ref{ccnn}). Similarly using the same logic and steps for the 
Pauli observable $Q=\vec{m}.\vec{\sigma}$, we obtain the following relation
\begin{equation}\label{infQ}
\Delta^2 Q(\rho_{AB})_{\inf}=\Delta^2 Q_A-\frac{CC(Q,Q)^2}{\Delta^2 Q_B}.
\end{equation}
Now adding Eqs. (\ref{infS}) and (\ref{infQ}) we obtain 
\begin{eqnarray}
&&\Delta^2 S(\rho_{AB})_{\inf}+\Delta^2 Q(\rho_{AB})_{\inf}= \Delta^2 S_A+\Delta^2 Q_A-\nonumber\\ &&\Big[\frac{CC(S,S)^2}{\Delta^2 S_B}+\frac{CC(Q,Q)^2}{\Delta^2 Q_B}\Big].\label{connected correlator}
\end{eqnarray}
It completes the proof.
\end{proof}
The numerators in the last two terms are the connected correlation functions. The denominators in the last two terms are local quantum uncertainties for the party $B$. 
Using the above equality, in the following, we state some equations that
capture the role of entanglement in the lower bound of uncertainty relations.
\subsubsection{Pure states}
\begin{proposition}
 For any two arbitrary Pauli observables $S=\vec{n}.\vec{\sigma}$ and $Q=\vec{m}.\vec{\sigma}$ and an arbitrary two-qubit pure state $|\psi_{AB}\rangle$, the 
following inequality holds
\begin{eqnarray}\label{con_uncertainty}
&&\Delta^2 S(|\psi_{AB}\rangle)_{\inf}+\Delta^2 Q(|\psi_{AB}\rangle)_{\inf}\geq\Delta^2 S_A+\Delta^2 Q_A-\nonumber\\ &&C^2(|\psi_{AB}\rangle)\Big[\frac{1}{\Delta^2 S_B}+\frac{1}{\Delta^2 Q_B}\Big],\label{concurrence_uncertainty}
\end{eqnarray}
where $C(|\psi_{AB}\rangle)$ denotes the concurrence of the state $|\psi_{AB}\rangle$.
\end{proposition}
\begin{proof}
Let us define $CC_{\max}=\max_{m,n}|CC(\sigma_m,\sigma_n)|$, which is the connected correlator maximized over all the observables acting on the parties $A$ and $B$ 
for a given quantum state. Thus $CC_{\max}\geq CC(Q,Q)$ for any particular $m$. Therefore from Eq. (\ref{connected correlator}), we get the following
\begin{eqnarray}\label{correlator_uncertainty}
&&\Delta^2 S(\rho_{AB})_{\inf}+\Delta^2 Q(\rho_{AB})_{\inf}\geq \Delta^2 S_A+\Delta^2 Q_A-\nonumber\\ &&CC_{\max}^2(\rho_{AB})\Big[\frac{1}{\Delta^2 S_B}+\frac{1}{\Delta^2 Q_B}\Big].
\end{eqnarray}
Using the expression in Eq. (\ref{concurrence and cc}), which states that for a pure two-qubit state the concurrence is given by the maximum connected correlator, i.e., 
$C(|\psi_{AB}\rangle)=CC_{\max}(|\psi_{AB}\rangle)$, we obtain the following
\begin{eqnarray}
&&\Delta^2 S(|\psi_{AB}\rangle)_{\inf}+\Delta^2 Q(|\psi_{AB}\rangle)_{\inf}\geq\Delta^2 S_A+\Delta^2 Q_A-\nonumber\\ &&C^2(|\psi_{AB}\rangle)\Big[\frac{1}{\Delta^2 S_B}+
\frac{1}{\Delta^2 Q_B}\Big],
\end{eqnarray}
which completes the proof.
\end{proof}

It is worthwhile to mention that the inequality derived above and other inequalities derived latter are always satisfied, unlike the inequality given in Eq. (\ref{reid_uncertainty}) which is valid for unsteerable states and a violation suggests that the state is entangled \cite{reid}.

Strikingly, the above equation directly shows how an entanglement monotone given by the concurrence is responsible for determining the lower bound of the inferred-variance
uncertainty relation for a pure two-qubit state. Some bounds are known for the uncertainties in party $A$ for two arbitrary Pauli observables and pure two-qubit states. Using that we can get
\begin{eqnarray}
&&\Delta^2 S(|\psi_{AB}\rangle)_{\inf}+\Delta^2 Q(|\psi_{AB}\rangle)_{\inf}\geq  1-|\vec{m}.\vec{n}|-\nonumber\\ 
&&C^2(|\psi_{AB}\rangle)\Big[\frac{1}{\Delta^2 S_B}+\frac{1}{\Delta^2 Q_B}\Big],
\end{eqnarray}
where we have used the relation $\Delta^2 S_A+\Delta^2 Q_A\geq 1-|\vec{m}.\vec{n}|$ \cite{Busch3}. 
However this is not a tight bound, and tighter state independent bounds
can be found in \cite{Abbott} for two arbitrary Pauli observables. 
\subsubsection{Mixed states}
\begin{proposition}\label{proposition G}
 For any two arbitrary Pauli observables $S=\vec{n}.\vec{\sigma}$ and $Q=\vec{m}.\vec{\sigma}$ and an arbitrary two-qubit state $\rho_{AB}$, the 
following inequality holds
\begin{eqnarray}\label{g_uncertainty}
&&\Delta^2 S(\rho_{AB})_{\inf}+\Delta^2 Q(\rho_{AB})_{\inf}\geq \Delta^2 S_A+\Delta^2 Q_A- \nonumber\\ && G(\rho_{AB})\Big[\frac{1}{\Delta^2 S_B}+\frac{1}{\Delta^2 Q_B}\Big], 
\end{eqnarray}
\end{proposition}
\begin{proof}
$G(\rho_{AB})$ has been shown to be an entanglement witness for two-qubit bipartite mixed states \cite{1,2,3}. Originally it was shown that it is given by the sum of the square of connected correlation functions for all Pauli observables as $G(\rho_{AB})=\sum_{i,j=1}^3 CC(\sigma_i,\sigma_j)^2$, 
which is nothing but 
$4\mbox{Tr}((\rho_{AB}-\rho_A\otimes\rho_B)^2)$.
As $G$ is invariant under any local unitary transformation \cite{1}, we have
$G(\rho_{AB})=\sum_{i,j=1}^3 CC(\sigma_i,\sigma_j)^2\geq CC(Q,Q)^2$ and as a result proposition \ref{proposition G}  follows.
\end{proof}
 Thus we are again able to connect the lower bound
of the inferred-variance uncertainty relation to an entanglement witness even for two-qubit mixed states and all observables. Note that for two-qubit pure states $G(|\psi_{AB}\rangle)=C^2(|\psi_{AB}\rangle)[2+C^2(|\psi_{AB}\rangle)]$\cite{1}. Therefore, Eq. (\ref{g_uncertainty}) does not reduce to Eq. (\ref{con_uncertainty}) for the pure two-qubit case and Eq. (\ref{con_uncertainty}) provides much better bound as $1<[2+C^2(|\psi_{AB}\rangle)]$. Furthermore, the entanglement detection given by $G(\rho_{AB})$ satisfies the inequality $G(\rho_{AB})\leq 1+2C^2(\rho_{AB})$ for any two-qubit bipartite mixed state \cite{1}, where $C(\rho_{AB})$ is
the concurrence of the mixed state $\rho_{AB}$. This relation can be used to get a bound such as 
\begin{eqnarray}
&\Delta^2 S(\rho_{AB})_{\inf}+\Delta^2 Q(\rho_{AB})_{\inf}\geq \Delta^2 S_A+\Delta^2 Q_A-\nonumber\\
&(1+2C^2(\rho_{AB}))\Big[\frac{1}{\Delta^2 S_B}+\frac{1}{\Delta^2 Q_B}\Big].
\end{eqnarray}

\subsection{Product inferred-variance uncertainty relations}
\subsubsection{Pure states}
\begin{proposition}\label{proposition product}
 For any two arbitrary Pauli observables $S=\vec{n}.\vec{\sigma}$ and $Q=\vec{m}.\vec{\sigma}$ and an arbitrary two-qubit pure state $|\psi_{AB}\rangle$, the 
following inequality holds
\begin{eqnarray}
\Delta^2 S(|\psi_{AB}\rangle)_{\inf}\Delta^2 Q(|\psi_{AB}\rangle)_{\inf}&\geq & \Delta^2 S_A\Delta^2 Q_A- \nonumber\\
&&\mkern-120mu      C^2(|\psi_{AB}\rangle)\Big[\frac{\Delta^2 S_A}{\Delta^2 S_B}+\frac{\Delta^2 Q_A}{\Delta^2 Q_B}\Big].\label{product_uncertainty_concurrence}
\end{eqnarray}
\end{proposition}
\begin{proof}
We start with the proof for a general density matrix, and the pure state case as above is derived as a special case of that.
Using Eqs. (\ref{infS}) and (\ref{infQ}) we find
\begin{eqnarray}
&\Delta^2 S(\rho_{AB})_{\inf}\Delta^2 Q(\rho_{AB})_{\inf}=\Delta^2 S_A\Delta^2 Q_A-\\
 &\Big[\frac{\Delta^2 S_A \, CC(S,S)^2}{\Delta^2 S_B}
+\frac{\Delta^2 Q_A \, CC(Q,Q)^2}{\Delta^2 Q_B}\Big]+\frac{CC(S,S)^2 \,CC(Q,Q)^2}{\Delta^2 S_B \, \Delta^2 Q_B}.\nonumber\label{product_inferred}
\end{eqnarray}
The last term is always greater or equals to zero. Therefore, we can write
\begin{eqnarray}
\Delta^2 S(\rho_{AB})_{\inf}\Delta^2 Q(\rho_{AB})_{\inf}&\geq & \Delta^2 S_A\Delta^2 Q_A-\nonumber\\
&&\mkern-120mu CC_{\max}^2(\rho_{AB})\Big[\frac{\Delta^2 S_A}{\Delta^2 S_B}+\frac{\Delta^2 Q_A}{\Delta^2 Q_B}\Big].
\end{eqnarray}
It is easy to see that for the case of pure states, since $CC_{\max}(|\psi_{AB}\rangle)=C(|\psi_{AB}\rangle)$, where $C(|\psi_{AB}\rangle)$ denotes the concurrence of the state 
$|\psi_{AB}\rangle$, we obtain the
bound as in proposition \ref{proposition product}.
\end{proof} 
\subsubsection{Mixed states}
Again, in the same vein as before, if we use the $G$ function
as the entanglement witness for mixed states, then we can write a similar equation for the product of variances for a mixed two-qubit state.
\begin{proposition} 
For any two arbitrary Pauli observables $S=\vec{n}.\vec{\sigma}$ and $Q=\vec{m}.\vec{\sigma}$ and an arbitrary two-qubit state $\rho_{AB}$, the 
following inequality holds
\begin{eqnarray}
\Delta^2 S(\rho_{AB})_{\inf}\Delta^2 Q(\rho_{AB})_{\inf}&\geq & \Delta^2 S_A\Delta^2 Q_A-\nonumber\\
&&\mkern-80mu G(\rho_{AB})\Big[\frac{\Delta^2 S_A}{\Delta^2 S_B}+\frac{\Delta^2 Q_A}{\Delta^2 Q_B}\Big].
\end{eqnarray}
\end{proposition}
The condition that gives a non-trivial bound for a mixed state is $\frac{1}{\frac{\Delta^2 S_A}{\Delta^2 S_B}+\frac{\Delta^2 Q_A}{\Delta^2 Q_B}}\geq G(\rho_{AB})\geq 1$.
We have highlighted how entanglement plays a part in affecting the value of the lower bound for the sum and product versions of 
inferred-variance uncertainty relations. In addition, the operators that maximize the connected correlator can be found out explicitly as given in \cite{adesso}, and as a result
one will be able to find an analytical expression for the above bounds in terms of the maximum connected correlation function for any two arbitrary Pauli observables. 
Furthermore, the connected correlation function can also be connected with some other correlation function such as mutual information, discord etc. Therefore, the lower bound of inferred-variance uncertainty relations can be written in terms of those correlation functions. We discuss this in the Appendix \ref{appendix mutual} and \ref{appendix discord}. We now move onto some examples.
\subsection{Examples}\label{example}
\subsubsection{Pure states}
Consider a pure two-qubit entangled state of the form $|\Psi\rangle=\cos\theta|00\rangle+\sin\theta|11\rangle$. If we choose $S=\sigma_x$ and $Q=\sigma_y$,
 we find
\begin{eqnarray}\label{example1}
&&\Delta^2 \sigma_x(|\Psi\rangle)_{\inf}+\Delta^2 \sigma_y(|\Psi\rangle)_{\inf}=2-2C^2(|\Psi\rangle),\nonumber\\
&&\Delta^2\sigma_x=\Delta^2\sigma_y=1,
\end{eqnarray} 
where $C(|\Psi\rangle)=\sin2\theta$ is the concurrence of the state $|\Psi\rangle$ \cite{concurrence}. Hence, for this state and observables we exactly saturate the uncertainty bound given in Eq. (\ref{concurrence_uncertainty}). As we have saturated the bound exactly, there is one nice operational advantage of the relation, which is as follows.
In an experiment we can measure $\Delta^2 \sigma_x(|\Psi\rangle)_{\inf}+\Delta^2 \sigma_y(|\Psi\rangle)_{\inf}$ and $\Delta^2\sigma_x+\Delta^2\sigma_y$. From there we can compute the
entanglement as 
\begin{eqnarray} \label{example2}
C^2(|\Psi\rangle)&=&\frac{1}{2}\Big((\Delta^2\sigma_x+\Delta^2\sigma_y)-\nonumber\\
&&[\Delta^2 \sigma_x(|\Psi\rangle)_{\inf}+\Delta^2 \sigma_y(|\Psi\rangle)_{\inf}]\Big).
\end{eqnarray}
Therefore, we can detect and quantify the entanglement present in a two-qubit pure state by using the inferred-variance uncertainty relation.
\begin{figure}[h]
\centering
\includegraphics[scale=0.65]{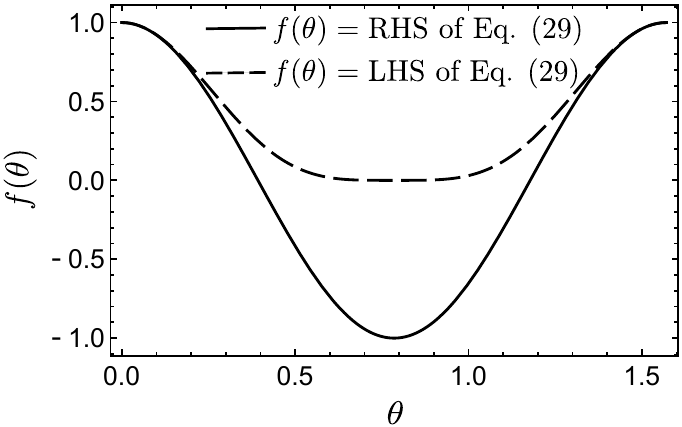}
\caption{Product of inferred variances for the observables $\sigma_x$, $\sigma_y$ and for the state $|\Psi\rangle$. The expressions of the solid and the dashed line correspond to the right hand side and the left hand side of Eq. (\ref{product_uncertainty_concurrence}) respectively.}
\label{qubit_product}
\end{figure}
For the case of product of inferred variances, we plot the left hand side and the right hand side of Eq. (\ref{product_uncertainty_concurrence}) with the state parameter $\theta$ in Fig. \ref{qubit_product}. 
The plot shows that the right hand side gives non-trivial bound for certain range of parameters of entangled pure states. It also shows that unlike the sum of inferred variances,
the bound for product of variances is not saturated for this class of pure states and for the observables $\sigma_x$, $\sigma_y$.
\subsubsection{Mixed states}
To check whether Proposition \ref{proposition G} gives us a non-trivial lower bound, we consider the Werner state of the form
\begin{equation}
\rho_W=p|\psi_s\rangle \langle \psi_s|+\frac{1-p}{4}I,
\end{equation}
where $|\psi_s\rangle=\frac{1}{\sqrt{2}}(|01\rangle - |10\rangle )$ is a singlet state. Furthermore, we consider the observable $S=\sigma_x$ and $Q=\sigma_y$. We find,
\begin{eqnarray}
&&\Delta^2 \sigma_x(\rho_W)_{\inf}+\Delta^2 \sigma_y(\rho_W)_{\inf}=2-2p^2,\nonumber\\
&&\Delta^2\sigma_x=1, \Delta^2\sigma_y=1, G({\rho_{W}})=3p^2.
\end{eqnarray} 
From here it is easy to verify that we can obtain a non-trivial lower bound for a range of $p$. To find the range, we note that the Werner state is entangled when $p> 1/3$ and the lower bound is greater than zero for $p<\frac{1}{\sqrt{3}}$. Using these two conditions we find $\frac{1}{\sqrt{3}}> p>\frac{1}{3}$, where a non-trivial lower bound is obtained.

\section{Local uncertainty relation violation and correlations}\label{local uncertainty}
In this section, we derive how the violation of local uncertainty relations can be quantified by an entanglement measure or witness. 
\begin{proposition}
We first note that for any two sets of arbitrary Hermitian operators $\{A_i\}$, $\{B_i\}$ and an arbitrary density matrix $\rho_{AB}$, the following
inequality holds
\begin{equation}\label{local uncertainty relation}
\sum_i\Delta^2 (A_i+B_i)\geq
L_A+L_B-
2|CC_{\max}| N,
\end{equation}
where $N$ represents the number of Hermitian operators in each set. 
\end{proposition}
\begin{proof}
Note that $\{A_i\}$ and $\{B_i\}$ act locally on the sub-systems $A$ and $B$ respectively. We start with 
\begin{eqnarray}
\Delta^2 (A_i+B_i)&=&\mbox{Tr}[(A_i^2\otimes I)\rho_{AB}]+\mbox{Tr}[(I\otimes B_i^2)\rho_{AB}]-\nonumber\\
&&\mbox{Tr}[(A_i\otimes I)\rho_{AB}]^2 -\mbox{Tr}[(I\otimes B_i)\rho_{AB}]^2+\nonumber\\
&&
2\mbox{Tr}[(A_i\otimes B_i)\rho_{AB}]-\nonumber\\
&&2\mbox{Tr}[(A_i\otimes I)\rho_{AB}]\mbox{Tr}[(I\otimes B_i)\rho_{AB}]\nonumber\\
&=&\Delta^2 (A_i)+\Delta^2 (B_i)+ 2\mbox{Tr}[(A_i\otimes B_i)\rho_{AB}]
\nonumber\\
&&-2\mbox{Tr}[A_i\rho_{A}]\mbox{Tr}[B_i\rho_{B}].
\end{eqnarray}
Using the expression $\sum_i\Delta^2 A_i +\sum_i\Delta^2 B_i\geq L_A+L_B$ as given in Eq. (\ref{local uncertainty relation violation}), we obtain
\begin{eqnarray}
\sum_i\Delta^2 (A_i+B_i)&\geq& L_A+L_B-2\sum_i \Big\{\langle A_i\rangle\langle B_i\rangle-\nonumber\\
&&\qquad\qquad\qquad\qquad\langle A_i\otimes B_i\rangle\Big\}\nonumber\\
&\geq& L_A+L_B-2|CC_{\max}| N,\label{loc uncert rela 1}
\end{eqnarray}
which completes the proof.
\end{proof}
In general, the above equation holds for an arbitrary bipartite quantum states. Again using the relation of connected correlator with concurrence for two-qubit pure state $|\psi_{AB}\rangle$ and putting that in Eq. \ref{loc uncert rela 1}, we get the following inequality
\begin{equation}
\sum_i\Delta^2 (A_i+B_i)\geq
L_A+L_B-
2C(|\psi_{AB}\rangle) N,
\end{equation}
where we consider that the Hermitian operators $\{A_i\}$ and $\{B_i\}$ are Pauli matrices in arbitrary directions, and $C(|\psi_{AB}\rangle)$ is the concurrence of the state $|\Psi\rangle$ \cite{concurrence}. Note that it might be possible to find similar relations in the continuous variable systems using the recent results for the local uncertainty relation violation in the continuous variable systems \cite{Marian2021}.

\section{Visibility of interference fringes}\label{visibility}
The uncertainty associated with a unitary operator $U$ in a given state $| \psi_{AB}\rangle$ can be expressed as 
$\Delta^2 U=\langle UU^\dagger\rangle-\langle U\rangle\langle U^\dagger\rangle=1-|\langle U\rangle|^2$, where $\langle U \rangle =\langle \psi| U|\psi\rangle$ \cite{massar}. Interestingly, it has a nice operational interpretation. 
The uncertainty of $U$ is intimately connected with the interference 
fringe visibility in a Mach-Zehnder 
interferometer \cite{sjoqvist}. Suppose a particle in a pure state $|\psi\rangle$ is fed to the interferometer and on one arm of the interferometer a unitary operator $U$ is applied,
then the fringe visibility of the interference pattern given by $\nu$ satisfies the relation $\Delta^2 U+\nu^2=1$, where $\nu=|\langle U\rangle|$. Earlier it was shown that using this property, the preparation 
uncertainty relation given by the unitary operators puts a non trivial constraint on the corresponding fringe visibilities \cite{bagchi}. Similarly we ask the  question - how is 
the constraint on fringe visibility modified when we have entangled quantum systems at our disposal ? In other words, can we see the effect of entanglement
directly in an interference experiments? We devise such an experiment in the next paragraph.

One of the interesting features of the equations given in \ref{local uncertainty} is that we can bound the visibility of interference fringes of the Mach-Zehnder interferometer 
for a complementary quantum state
by the correlation content in a bipartite quantum state. By complementary quantum state, we mean that if we have a bipartite quantum system $\rho_{AB}$, 
then $\rho_A$ and $\rho_B$ are
the complementary quantum states.
For this, we consider the Pauli operators that are both unitary and Hermitian.
We know that the visibility of an interference pattern in a Mach-Zehnder
interferometer shares a complementarity relation with the uncertainty of the unitary operator which is affecting one of the arms of the interferometer.
We note that the uncertainty based on the variance of a unitary operator $U$ is 
quantified like $\Delta^2 U=1-|\mbox{Tr}(U\rho_{AB})|^2$. Taking this and considering an arbitrary pure two-qubit state $|\psi_{AB}\rangle$, it is straightforward to extend the local uncertainty 
violation for the unitary operators $U$ and $V$ as follows
\begin{equation}
\nu_U^2+\nu_V^2\geq
2[1-C(|\psi_{AB}\rangle)]-\Delta (U+V)^2, 
\end{equation}
where $\nu$ represents the fringe visibility, the two terms on the left-hand side are calculated for the two complementary quantum states, and the last term on the 
right-hand side
is calculated on the bipartite quantum system.
The above equation directly shows that when we take a bipartite state and calculate the corresponding uncertainties, then the sum of the interference fringe 
visibilities obeys a 
bound dependent on the concurrence of the system.
\section{Applications of inferred-variance uncertainty relations} \label{application}
Apart from the fundamental importance of inferred-variance uncertainty relations, we point out a few applications of these relations in a few quantum information protocols, such as witnessing entanglement, measuring entanglement, analysis of the security of the quantum cryptographic protocols, etc. However, to implement these relations in practice, the left-hand side of the inferred-variance uncertainty relations, involving inferred variance, needs to be measured in an experiment. However, it can be as straightforward as
a Bell experiment. As we saw, one can compute the inferred variance using the joint probability distribution for different measurement outcomes. The same object is measured
in a Bell experiment. Moreover, in a recent experiment \cite{experiment}, the entropic uncertainty relation by Berta {\it et al.} has been verified. Using this same experimental technique, one may be able to measure the left-hand side of the inferred-variance uncertainty relations.

\subsection{Entanglement Measurement}
      Using a specific set of observables, we can measure the entanglement of a bipartite
      pure state. If Alice and Bob share a partially entangled
      pure state, then by measuring the observables $\sigma_x$ and
      $\sigma_y$, they can determine the concurrence of the state, using the result given in Eq. (\ref{example2}).

\subsection{Entanglement witness}
As the lower bound of inferred-variance uncertainty relations depends on the entanglement, we can construct entanglement witnesses. For a pure two-qubit state, we can put a lower bound on the concurrence of the state using {\em arbitrary} observables. Rewriting Eq. (\ref{concurrence_uncertainty}) as
\begin{equation}
C^2 \geq \mathcal{C}_A,
\end{equation}
where 
\begin{equation}\label{lower1}
\mathcal{C}_A=\frac{\Delta^2 S_B \Delta^2 Q_B}{\Delta^2 S_B+\Delta^2 Q_B}\Big[\Delta^2 S_A+\Delta^2 Q_A-
\Delta^2 S_{\inf}^A-\Delta^2 Q_{\inf}^A\Big].
\end{equation}
The inequality in Eq. (\ref{concurrence_uncertainty}) is derived for the inferred variance on Alice's side depending on Bob's outcome. Similar inequality can be obtained for the case where inferred variances are calculated for Bob's side depending on Alice's outcome. In that case the lower bound will be
\begin{equation}\label{lower2}
\mathcal{C}_B=\frac{\Delta^2 S_A \Delta^2 Q_A}{\Delta^2 S_A+\Delta^2 Q_A}\Big[\Delta^2 S_B+\Delta^2 Q_B-
\Delta^2 S_{\inf}^B-\Delta^2 Q_{\inf}^B\Big].
\end{equation}
To avoid confusion between inferred variances on Alice's side and Bob's side, we have added a superscript $A$ and $B$ to represent inferred variances on Alice's and Bob's side respectively. Combining $\mathcal{C}_A$ and $\mathcal{C}_B$ we can find a lower bound for concurrence as following
\begin{equation}
C^2\geq\max{\{\mathcal{C}_A,\mathcal{C_B}\}},
\end{equation}
where a nonzero lower value ensures the presence of entanglement in the state.  As discussed above, for some particular measurements this bound can be saturated. Therefore, {\em the uncertainty relations can also be used to measure the entanglement of pure states.} Similarly in case of an arbitrary two-qubit state we can put a lower bound on $G$ as well using Proposition \ref{proposition G} as follows
\begin{equation} 
G \geq \max{\{\mathcal{C}_A,\mathcal{C_B}\}},
\end{equation}
where $G\geq 1$ implies that the state is entangled and $\mathcal{C}_A$ and $\mathcal{C}_B$ are calculated for a general two-qubit state $\rho_{AB}$.
In this way, by measuring the inferred variances in an experiment, we can detect, and even measure, the entanglement in the state. 
\subsection{QKD Protocol}
One can use inferred-variance uncertainty relations, in particular the one given in equation
(\ref{example1}) for a pure state with a specific measurement choice, for secure key generation. 
For a maximally entangled pure state this relation is
\begin{equation}\label{qkdprotocol}
\Delta^2 \sigma_x(|\Psi\rangle)_{\inf}+\Delta^2 \sigma_y(|\Psi\rangle)_{\inf}=0.
\end{equation} 
In this quantum key distribution protocol, similar to the protocol of Ekert \cite{ekert}, Alice and Bob  have one qubit each of an entangled pair in the Bell state $|\phi^+\rangle=\left(|00\rangle+|11\rangle\right)/\sqrt{2}$. Alice and Bob both measure their respective particle in the bases of $\sigma_x$ $\left\{\left(|0\rangle+|1\rangle\right)/\sqrt{2}, \left(|0\rangle-|1\rangle\right)/\sqrt{2}\right\}$ or $\sigma_y$ $\left\{\left(|0\rangle+i|1\rangle\right)/\sqrt{2}, \left(|0\rangle-i|1\rangle\right)/\sqrt{2}\right\}$. Then, as in Ekert protocol \cite{ekert}, one can measure the joint probability distributions, and thus inferred variance of the observables. Then violation of (\ref{qkdprotocol}) would indicate the presence of an intruder, Eve. Alice and Bob can also establish a secret key like in Ekert protocol, when they have correlated measurement outcomes. The advantage of our protocol over Ekert's protocol is key rate. In the case Ekert protocol, Alice and Bob
make three measurements each so as to establish a key and observe
Bell violation, and so key rate is ${2 \over 9}$.
In our protocol, only two measurements each are required for
both establishing a key and security, so key rate is ${1 \over 2}$.
However, there exit BBM protocol \cite{bbm} where like our
protocol, only two measurements by each party are needed.

Furthermore, not only in discrete systems but also these uncertainty relations may be useful in continuous variable systems. 
There are various quantum cryptographic protocols that use conditional variances to analyse the security of their protocols \cite{soh, grosshans, weedbrook}.
As shown in these references, the conditional variance usage in finding the secret key rate in a cryptographic protocol is a routine practice as the expression of the secret key rate is intimately linked with the expression of the conditional variance. Thus the bounds on conditional variance bound the secret key rate. The bound given by our uncertainty relations may prove to be useful in the security analysis of such cryptographic protocols. However, a full analysis of it is beyond the scope of this paper.
\section{Conclusions and Future Directions}\label{conclusion}
In this paper, we have formulated inferred-variance uncertainty relations in the presence of quantum entanglement. We show that the lower bounds for inferred-variance uncertainty relations can be linked with the entanglement measure and witness such as the concurrence for two-qubit pure states and the $G$ function for two-qubit mixed states. These uncertainty relations can have several
applications.  We have shown that inferred-variance uncertainty relations are useful to detect entanglement in a two-qubit state. In the case of a pure state, one can use these relations to measure the entanglement in the state. These relations can also 
be useful to generate a secure cryptographic key. We have given a 
QKD protocol that has higher key rate than Ekert protocol and uses
the derived uncertainty relations to detect a breach of security.

In addition, we show how the connected correlation function plays an important role in determining the lower bound of inferred-variance uncertainty relations. 
 Thus the relation between the connected correlation function and other measures/witnesses of entanglement is an important
 direction for further research. 
One can also go beyond a bipartite situation. In the case of genuine multipartite correlations, the multipartite connected correlator comes into play which cannot be derived from the bipartite connected correlation function. However, perhaps uncertainty relations in this case cannot be obtained if we only consider variance. Therefore for this purpose, one may need a quantifier 
which is of higher order than the variance. Besides, there has been some research on the 
genuine multipartite steering criteria using inferred uncertainty relations, and it will be 
interesting to see if the genuine multipartite entanglement plays a role in such situations. 

\section*{ACKNOWLEDGEMENTS}

S. B. acknowledges the support provided by 
PBC Post-Doctoral Fellowship at Tel Aviv University. S. B. also thanks the Institute of Physics, Bhubaneswar, India for their hospitality. 
C. D. acknowledges the support from the ``Quantum Optical
Technologies” project, carried out within the International Research Agendas programme of the Foundation for Polish Science co-financed by the European Union
under the European Regional Development Fund. P. A. acknowledges the 
support form Department of Science and Technology, India,  through the
project DST/ICPS/QuST/Theme-1/2019.
\\
\\
\appendix

\section{Expression of inferred variance}\label{expression of inferred}
 The inferred variance of the observable $S=\vec{n}.\vec{\sigma}$ in state $\rho_{AB}$ is given by the following expression. 
\begin{widetext}
\begin{eqnarray}\nonumber
    \Delta^2 S(\rho_{AB})_{\inf}=\frac{1}{4}\Bigg[\Bigg(\frac{-(1-\vec{n}.\vec{s}-\vec{n}.\vec{r}+n_1^2t_{11}+n_2^2t_{22}+n_3^2t_{33}+n_1n_2(t_{12}+t_{21})+n_1n_3(t_{13}+t_{31})+n_3n_2(t_{32}+t_{23}))^2}{(1-\vec{n}.\vec{s})}\times\nonumber\\ \frac{(-1+\vec{n}.\vec{s}-\vec{n}.\vec{r}+n_1^2t_{11}+n_2^2t_{22}+n_3^2t_{33}+n_1n_2(t_{12}+t_{21})+n_1n_3(t_{13}+t_{31})+n_3n_2(t_{32}+t_{23}))}{(1-\vec{n}.\vec{s})}\Bigg) \nonumber\\+\Bigg(\frac{(1-\vec{n}.\vec{s}-\vec{n}.\vec{r}+n_1^2t_{11}+n_2^2t_{22}+n_3^2t_{33}+n_1n_2(t_{12}+t_{21})+n_1n_3(t_{13}+t_{31})+n_3n_2(t_{32}+t_{23}))}{(1-\vec{n}.\vec{s})}\times\nonumber\\ \frac{(-1+\vec{n}.\vec{s}-\vec{n}.\vec{r}+n_1^2t_{11}+n_2^2t_{22}+n_3^2t_{33}+n_1n_2(t_{12}+t_{21})+n_1n_3(t_{13}+t_{31})+n_3n_2(t_{32}+t_{23}))^2}{(1-\vec{n}.\vec{s})}\Bigg)\nonumber\\ +\Bigg(\frac{(-1-\vec{n}.\vec{s}+\vec{n}.\vec{r}+n_1^2t_{11}+n_2^2t_{22}+n_3^2t_{33}+n_1n_2(t_{12}+t_{21})+n_1n_3(t_{13}+t_{31})+n_3n_2(t_{32}+t_{23}))^2}{(1+\vec{n}.\vec{s})}\times\nonumber\\ \frac{(1+\vec{n}.\vec{s}+\vec{n}.\vec{r}+n_1^2t_{11}+n_2^2t_{22}+n_3^2t_{33}+n_1n_2(t_{12}+t_{21})+n_1n_3(t_{13}+t_{31})+n_3n_2(t_{32}+t_{23}))}{(1+\vec{n}.\vec{s})}\Bigg) \nonumber\\-\Bigg(\frac{(-1-\vec{n}.\vec{s}+\vec{n}.\vec{r}+n_1^2t_{11}+n_2^2t_{22}+n_3^2t_{33}+n_1n_2(t_{12}+t_{21})+n_1n_3(t_{13}+t_{31})+n_3n_2(t_{32}+t_{23}))}{(1+\vec{n}.\vec{s})}\times\nonumber\\ \frac{(1+\vec{n}.\vec{s}+\vec{n}.\vec{r}+n_1^2t_{11}+n_2^2t_{22}+n_3^2t_{33}+n_1n_2(t_{12}+t_{21})+n_1n_3(t_{13}+t_{31})+n_3n_2(t_{32}+t_{23}))^2}{(1+\vec{n}.\vec{s})}\Bigg)\Bigg]\nonumber.
\end{eqnarray}
\end{widetext}
This expression can be rearranged to give the final expression written in Eq. (\ref{inferred varianceS}) of the main text.

\section{Inferred-variance uncertainty relations with some other correlation functions}\label{appendix mutual}
Here we outline some other correlation functions which can be used in the lower bound of inferred-variance uncertainty relations. There is a measure of entanglement for pure states called the covariance entanglement measure, defined as follows.
If we have operators $A^{(1)}=A\otimes I$ and
$B^{(2)}=I\otimes B$, then the covariance entanglement is defined as the following
\begin{equation}
~~~~~~~~E_{\mbox{cov}}(|\psi_{AB}\rangle)=\max_{U=U^{(1)}\otimes U^{(2)}}\mbox{Cov}_{U\rho U^\dagger}(A^{(1)},B^{(2)}),
\end{equation}
where, $\mbox{Cov}(A^{(1)},B^{(2)})=\mbox{Tr}(A^{(1)}B^{(2)}\rho_{AB})-\mbox{Tr}(A^{(1)}\rho_{AB})\mbox{Tr}(B^{(2)}\rho_{AB})$ and the maximization over the unitary matrices clearly represents the invariance under local unitary transformations.
It is easy to verify that this quantity is nothing but 
the connected correlation function of the operators $A$ and $B$ as mentioned above in the context of defining $A^{(1)}$ and $B^{(2)}$. 

However, the covariance entanglement measure is
not a good entanglement measure for mixed states \cite{Davis}. Not only
the entanglement correlation functions, but the total correlation in a quantum state also satisfies an important inequality with the connected correlator. The quantum mutual information which quantifies both the quantum and classical correlation satisfies the following inequality \cite{mi1}
with the connected correlator
\begin{equation}\label{mutual information}
I(A:B)\geq \frac{CC(M_A,M_B)^2}{2||M_A||^2||M_B||^2},
\end{equation}
where $M_A$ and $M_B$ represent two observables in Hilbert spaces $\mathbb{C}^A$ and $\mathbb{C}^B$ respectively and the mutual information is defined as $I(A:B)=H(\rho_A)+H(\rho_B)-H(\rho_{AB})$ for a state $\rho_{AB}$, where $H(\rho)$ represents the von-Neumann entropy of a state $\rho$.
All the above relations help us to bring the correlation measures inside the conditional variance uncertainty relations for the correlated quantum systems. 

Now replacing $CC_{\max}$ by the $E_{\mbox{cov}}$ in Eq. (\ref{correlator_uncertainty}) for a pure state $|\psi_{AB}\rangle$, we obtain 
\begin{eqnarray}
&&\Delta^2 S(|\psi_{AB}\rangle)_{\inf}+\Delta^2 Q(|\psi_{AB}\rangle)_{\inf}\geq \Delta^2 S_A+\Delta^2 Q_A-\nonumber\\ 
&&E^2_{\mbox{cov}}(|\psi_{AB}\rangle)\Big[\frac{1}{\Delta^2 S_B}+\frac{1}{\Delta^2 Q_B}\Big]. 
\end{eqnarray}
Again using Eq. (\ref{mutual information}), we can obtain a lower bound that involves the quantum mutual information \cite{mi1} for the case of mixed states. This is given by 
\begin{eqnarray}
&\Delta^2 S(\rho_{AB})_{\inf}+\Delta^2 Q(\rho_{AB})_{\inf}\geq\Delta^2 S_A+\Delta^2 Q_A-\nonumber\\
& 2I(A:B)\bigg[\frac{||(\vec{n}.\vec{\sigma})||^4}{\Delta S_B^2}+\frac{||(\vec{m}.\vec{\sigma})||^4}{\Delta Q_B^2}\bigg].
\end{eqnarray}
\section{inferred-variance uncertainty relations and quantum discord}\label{appendix discord}
The quantum discord was proposed to capture the 
quantum correlations that cannot be captured by the entanglement. 
Along this vein, for a bipartite quantum state, a discord-like measure of quantum correlations was introduced by Girolami et al \cite{girolami}. It is based on Skew Information. This measure was shown to have features of discord like measures of quantum correlations. We briefly discuss this in the next section.

There are several ways to quantify the uncertainty on a measurement. Entropic quantities or the variance have been employed in many ways as indicators of uncertainty. One such measure is also the Skew Information, also employed to study the uncertainty relations. Skew information is defined as 
\begin{equation}\mathcal{I}(\rho,K)=-1/2\mbox{Tr}\{[\sqrt{\rho},K]^2\}.
\end{equation}
 Here $K$ is some local observable and $\rho$ is a bipartite state. As a property, we have that 
$\mathcal{I}(\rho,K)$ is always smaller than the variance, with equality for pure states \cite{girolami}. 

The local quantum uncertainty has been defined as the minimum skew information attainable with a single measurement on one of the local parties. As stated earlier it has been shown to satisfy all the known bona fide criteria for a discord like quantifier of quantum correlations. It has a very nice geometrical
interpretation of quantum discord \cite{girolami} as follows.The local quantum uncertainty in a general state $\rho_{AB}$ of a $\mathbb{C}^2\otimes\mathbb{C}^d$ system, can be reinterpreted geometrically as the minimum squared Hellinger distance between $\rho_{AB}$  and the state after a least disturbing root-of-unity local unitary operation applied on the qubit $A$, in a spirit close to that adopted to define ‘geometric discords’ based on other metrics. It is shown as follows. 

The squared Hellinger distance between two states is defined as $D^2(\rho,\chi)=\frac{1}{2}\mathrm{Tr}((\sqrt{\rho}-\sqrt{\chi})^2)$. For $K_A=\vec{n}.\vec{\sigma}$, the $I(\rho,K_A)=1-\mathrm{Tr}(\rho^{\frac{1}{2}}K_A\rho^{\frac{1}{2}}K_A)=1-\mathrm{Tr}(\rho^{\frac{1}{2}}(K_A\rho K_A)^{\frac{1}{2}})=D(\rho,K_A) $. Therefore, minimizing over the local observables $K_A$ gives one the geometric interpretation of the local quantum uncertainty in terms of the Hellinger distance for $\mathbb{C}^2\otimes\mathbb{C}^d$ quantum systems.

Here we prove a proposition that relates inferred variance to this definition of quantum discord.

\begin{proposition}
For any two arbitrary Pauli observables $S=\vec{n}.\vec{\sigma}$ and $Q=\vec{m}.\vec{\sigma}$ and an arbitrary two-qubit density matrix $\rho_{AB}$, the 
following inequality holds
\begin{eqnarray}\label{discord_uncertainty}
&&\Delta^2 S(\rho_{AB})_{\inf}+\Delta^2 Q(\rho_{AB})_{\inf}\geq \Delta^2 S_A+\Delta^2 Q_A- \nonumber\\ &&\frac{1}{D}\big[CC(Q,Q)^2+CC(S,S)^2\big].
\end{eqnarray}
\end{proposition}

\begin{proof}
The denominator in Eq. (\ref{connected correlator}) is an expression of local quantum uncertainty which is a signature of the quantum correlations in bipartite quantum
systems \cite{girolami}. 
As mentioned in the section on background, the variance is always greater than the skew information 
which is a better quantifier of the local quantum uncertainty and can be interpreted as the quantum discord, when the measurement is applied to party $B$. From this we get that 
$\frac{-1}{\Delta^2 S_B}\geq \frac{-1}{\mathcal{I}(\rho_{AB},S)}$. Putting this relation in Eq. (\ref{connected correlator}) and noting that we interpret $\mathcal{I}(\rho_{AB}, S)$ as a measure of quantum correlations for the quantum discord type in $\mathbb{C}^2 \otimes \mathbb{C}^d$ quantum systems and denoting it by $D$, we obtain the following
\begin{eqnarray}
&&\Delta^2 S(\rho_{AB})_{\inf}+\Delta^2 Q(\rho_{AB})_{\inf}\geq \Delta^2 S_A+\Delta^2 Q_A- \nonumber\\ &&\frac{1}{D}\big[CC(Q,Q)^2+CC(S,S)^2\big].
\end{eqnarray}
\end{proof}

In the above equation, we do not replace the connected correlation functions with the maximized value, as this further weakens the bound.  Here we discuss in brief the similarities with the bounds in \cite{berta, pati} and here. The bound given in \cite{berta} shows that the presence of entanglement lowers
the uncertainty bound, however in \cite{pati}, the quantum discord can tighten the bound by increasing it in certain cases, 
where the measurement is performed on party $A$. Here also in the case 
of inferred-variance uncertainty relations, the connected correlator acts as an entanglement signature for pure states
and it brings down the lower bound. Whereas the discord as quantified by local quantum uncertainty acts in a different way. Greater is the value of $D$, 
greater is the value of $-\frac{1}{D}$, as a result it therefore acts in a similar way as in case of \cite{pati}, i.e., greater value of the quantum discord helps 
to increase the value of the lower bound. However, the point of difference is that the measurement is performed
on party $B$. Hence, entanglement and discord behave differently in determining the lower bound. 

\begin{figure}[h]
\centering
\includegraphics[scale=0.55]{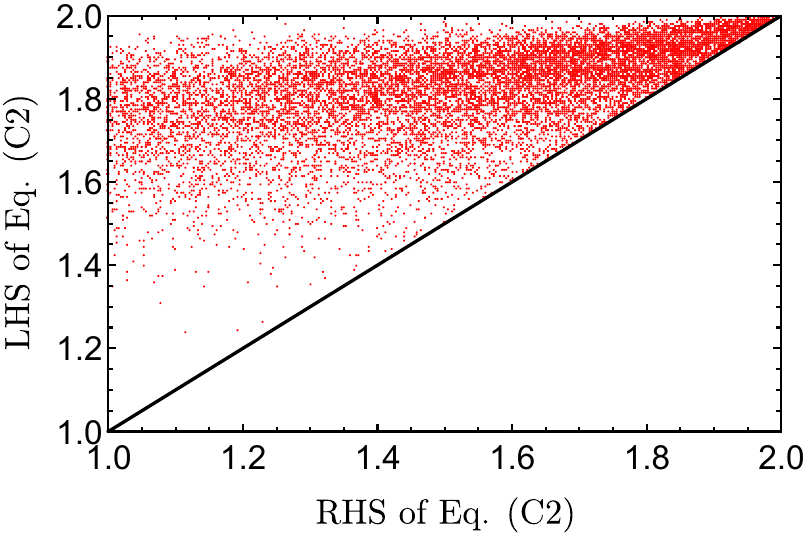}
\caption{The x and the y axis represent the right hand side and the left hand side of Eq. (\ref{discord_uncertainty}) respectively. The dots (dark grey, red online) represent randomly generated states.  }
\label{qubit_discord}
\end{figure}

We check that for various mixed random quantum states, and see that for quite some states we get a non trivial lower bound for 
the inequality given in Eq. (\ref{discord_uncertainty}) as shown in Fig. \ref{qubit_discord}. The x-axis and y-axis in Fig. \ref{qubit_discord} represent the right hand side 
and the left hand side of Eq. (\ref{discord_uncertainty}) respectively. The diagonal black line is provided to give the reference that left hand side is always greater than or equal to the 
right hand side of Eq. (\ref{discord_uncertainty}). 

\end{document}